\documentclass[12pt]{article}

\usepackage{graphicx}
\usepackage{bm}
\usepackage{amsmath}
\usepackage{amssymb}

\def\be{\begin{equation}}
\def\ee{\end{equation}}
\def\bea{\begin{eqnarray}}
\def\eea{\end{eqnarray}}

\def\ie{{\it i.e.}, }

\begin{document}

\begin{flushright}
\end{flushright}

\pagestyle{plain}

\begin{center}
\vspace{2.5cm} {\Large {\bf Coordinate/Field Affinity: \\ \vspace{0.5cm} A Proposal For Confinement}}

\vspace{1cm}

Amir H. Fatollahi

\vspace{.5cm}

{\it Department of Physics, Alzahra University, \\ P. O. Box 19938, Tehran 91167, Iran}

\vspace{.3cm}

\texttt{ahfatol@gmail.com}

\vskip .5 cm
\end{center}

\begin{abstract}
It is argued that demanding for similar characters between the coordinates of space-time
and the fields would sound that non-Abelian gauge theories might be formulated
most naturally based on fields depending on matrix coordinates.
\end{abstract}

\vspace{3cm}


\newpage
According to one possible interpretation of the special relativity agenda, it would be
meaningful to demand for similar characters between the coordinates of space-time and the
fields living in it. In particular, the space-time coordinates as well as the electromagnetic
potentials should transform equivalently, hence expecting mixing between time and space
under boost transformations. Also by this way of interpretation, the super-space
formulation of supersymmetric field and string theories is just a logical continuation
of the special relativity program: inclusion of anti-commutating coordinates as
representatives of the fermionic degrees of freedom of the theory.

In \cite{fat-1}-\cite{fat-6} a model was considered based on the possibility
that matrix coordinates can be used for reproducing the features we know about
non-Abelian gauge theories. The model was originated by the D0-branes
\cite{9510017,tasi} of string theory, for which it is known that their degrees of freedom
are captured by matrices, rather than numbers \cite{9510135}. The concerned model
has shown its ability to reproduce or cover some features and expectations in hadron
physics \cite{fat-1}-\cite{fat-3}. Some of these features and expectations are:
phenomenological inter-quark potentials, the behavior of total scattering amplitudes,
rich polology of scattering amplitude, behavior in large-$N$ limit, and the whiteness of
bound states with respect to the SU($N$) sector of the external fields.

The aim of this work is to reconsider the possible relevance of matrix coordinates
based on the proposal that space-time coordinates and fields living
in it should have similar characters. In this way we deal with the situation in which
the fields have extra indices coming from the matrix character of them.
In particular we consider the case with non-Abelian gauge theories, assuming that
the extra structure of fields should come with the coordinates too.
It is argued that the assumed similar characters, together with gauge symmetry
considerations, lead us to suggest that in the regime that symmetry transformation
would mix different components of fields, \ie the non-zero gauge coupling regime,
we should expect that the dynamics of theory is most naturally formulated based
on fields depending on matrix coordinates. As a consequence, each gauge sector contributes
equivalently in the building blocks of the theory.

The dynamics of a charged particle in presence of an electric field is given by
\bea\label{1}
m\,\ddot{\bm{x}}= e\, \bm{E}
\eea
in which $e$ represents the strength of coupling. Now we assume that the electric field
has some extra indices, those coming from the so-called internal-space. Demanding for
similar characters for fields and coordinates, one might assume that the same indices
should come with the coordinates too, translating the fact that particles charged under
different fields should satisfy different equations. In this way one simply writes:
\bea\label{2}
m\, \ddot{\bm{x}}^{a} = g\, \bm{E}^{\,a},\;\;\;\;\;\;\;a=0,1,2,\cdots,M,
\eea
in which $g$ is representing the coupling strength in the new theory. Further, as quite often,
we might require that the theory would be invariant under kinds of linear transformations of fields
in internal-space, namely
\bea\label{3}
\bm{E}^{\,a} \to \bm{E}\,'^{\,a}=\Lambda^{a}_{\;\; b} \,\bm{E}^{\,b}
\eea
For the moment $\Lambda^a_{\;\;b}$'s are globally defined; they are constants.
Requiring that the dynamics is equivalently formulated by either $\bm{E}^{\,a}$ or
$\bm{E}\,'^{\,a}$ forces us to demand that (\ref{2}) should transform
in a covariant way under the linear transformations,
\bea\label{3.05}
\bm{x}^{\,a} \to \bm{x}'^{\,a}=\Lambda^{a}_{\;\; b} \,\bm{x}^{\,b}
\eea
As a consequence each degree of freedom is mixed with others in an indistinguishable
and inseparable way. In other words, as none of $\bm{x}^{a}$'s is individually measurable,
the problem at hand should always be considered for a ``collection" consisting
all degrees of freedom.

One important issue would be the coordinate dependence of fields as well as the
transformation parameters. As mentioned we always have to deal with, rather than
a single, a collection of coordinates, and the purpose is to introduce the dependence
on coordinates in a way compatible with transformations. Fortunately, at least for
the most interested cases, there is a solution for the problem of dependence.
Assume that the transformation (\ref{3}) is given near identity by:
\bea\label{3.1}
\Lambda^{a}_{\;\; b} = \delta^{\,a}_{\;\;b} + \,f^{\,ac}_{\;\;\;\;\;b}\,\Lambda_{c}
\eea
in which $f^{\,ab}_{\;\;\;\;\;c}$'s are constants, and $\Lambda_{c}$'s are new parameterization
representing the transformation. In this case the transformation generators $T^{\,a}$'s satisfy
the commutation relation
\bea\label{4}
\big[\, T^{\,a},T^{\,b}\big]= {\rm i} f^{\,ab}_{\;\;\;\;\;c}\,T^{\,c}.
\eea
By this set of generators we introduce
\bea\label{5}
\mathbf{E}:=\bm{E}_{\,a}\, T^{\,a},\;\;\;\mathbf{x}:=\bm{x}_a\,T^{\,a}.
\eea
for which we have by (\ref{2})
\bea\label{6}
m\, \ddot{\mathbf{x}} = g\, \mathbf{E}.
\eea
In this case the transformation of $\bm{E}^{\,a}$'s can be given in terms of $\mathbf{E}$ by
\bea\label{7}
\mathbf{E}\to \mathbf{E}\,'= U \, \mathbf{E}\, U^{-1}.
\eea
in which $U$ is a proper invertible matrix.
Now we can present the solution for dependence problem. First let us consider the case
with global transformation, that is $\Lambda^{a}_{\;\;b}$'s, and so $U$,
do not depend on $\bm{x}^a$'s and $t$. By (\ref{6}) we simply expect
\bea\label{8}
\mathbf{x}\to \mathbf{x}'= U \, \mathbf{x}\, U^{-1}.
\eea
and so it would be enough if the dependence of $\bm{E}^{\,a}$'s is given by
dependence of $\mathbf{E}$ on $\mathbf{x}$, that is to assume
\bea\label{9}
\mathbf{E}=\mathbf{E}(\mathbf{x}).
\eea
In right-hand-side of above we simply mean a formal expansion in powers of $\mathbf{x}_i$'s,
the spatial components of $\mathbf{x}$. One example of such expansions may be assumed as
\bea\label{10}
\mathbf{E}_i = \mathbf{E}^{(0)}_i + \mathbf{E}^{(1)}_{ij} \, \mathbf{x}^j
+ \mathbf{E}^{(2)}_{ijk} \,  \mathbf{x}^j\,\mathbf{x}^k+ \cdots
\eea
As we mention, here one encounters with the ordering ambiguity for the $\mathbf{x}_i$'s
appearing in expansion. Henceforth we hire the symmetrization prescription, by which in
each term all possible permutations of $\mathbf{x}_i$'s equally contribute.
It is quite obvious that the transformations (\ref{7}) and (\ref{8}), together with the equation
of motion (\ref{6}) are consistent, at least for the case with global $U$.
The case with local transformation needs more modifications. Here our experience with gauge symmetry
can lead us to a resolution. We introduce the one dimensional gauge potential $a_0(t)$, by which
we have the one dimensional covariant derivative
\bea\label{10.1}
D_0=\frac{{\rm d}}{{\rm d}t} - {\rm i}\, a_0
\eea
The transformation laws now are given by
\bea\label{11}
\mathbf{x}&\to& \mathbf{x}'= U \, \mathbf{x}\, U^{-1},\nonumber\\
a_0(t) &\to&  a'_0(t)=U\, a_0(t)\, U^{-1} + {\rm i}\, U\, \frac{{\rm d}}{{\rm d}t}\, U^{-1},
\eea
in which $U(\mathbf{x},t)=\exp\big({\rm i} \Lambda(\mathbf{x},t)\big)$, and $\Lambda(\mathbf{x}_i,t)$
depends on $\mathbf{x}$ in the same way described for $\mathbf{E}(\mathbf{x})$.
We note that though $U(\mathbf{x},t)$ depends on $\mathbf{x}$, due to the total derivative
$\displaystyle{\frac{{\rm d}}{{\rm d}t}}$, $a'_0(t)$ still depends only on time. By
\bea\label{11.1}
D_0\, \mathbf{x} = \dot{\mathbf{x}} - {\rm i}\, [a_0,\mathbf{x}],
\eea
we have the following
\bea\label{12}
D_0 \,\mathbf{x} &\to& D_0\,'\, \mathbf{x}'= U\,\, D_0\, \mathbf{x}\,\, U^{-1}, \nonumber\\
D_0 D_0 \,\mathbf{x} &\to& D_0\,' D_0\,' \,\mathbf{x}'=U\,\, D_0D_0\, \mathbf{x}\,\, U^{-1}.
\eea
By these all we modify the equation of motion (\ref{6}) as follows
\bea\label{13}
m\, D_0 D_0 \,\mathbf{x} = g\, \mathbf{E}(\mathbf{x},t).
\eea
and obviously it transforms in a covariant way under the transformation laws (\ref{11}).
We may propose an action by which we can derive the equation of motion. Having in mind that
magnetic-like terms should also be included one may propose
\bea\label{14}
S[a_0,\mathbf{x}]=\int {\rm d}t \;{\rm Tr}\, \bigg(\frac{1}{2}\, m\, D_0\,\mathbf{x}\cdot D_0\,\mathbf{x} - g \, \mathrm{A}_0(\mathbf{x},t)
+ g\, D_0\,\mathbf{x}\cdot \mathbf{A}(\mathbf{x},t) \bigg),
\eea
in which $\big(\mathrm{A}_0(\mathbf{x},t),\mathbf{A}(\mathbf{x},t)\big)$ play the role of the potentials.
Then the equations of motion for $\mathbf{x}$ and $a_0$ can be derived straightforwardly,
\bea\label{15}
&~&mD_0D_0\, \mathbf{x}_i=g\, \big(\mathbf{E}_i(\mathbf{x},t) + \underbrace{D_0\,\mathbf{x}^j\,\mathbf{B}_{ji}(\mathbf{x},t)}\big),\\
&~&m[\mathbf{x}_i,D_0\,\mathbf{x}^i]=g\,[\mathbf{A}_i(\mathbf{x},t),\mathbf{x}^i],
\eea
in which
\bea\label{16}
\mathbf{E}_i(\mathbf{x},t)&:=&-\delta_i \mathrm{A}_0(\mathbf{x},t)-\partial_t \mathbf{A}_i(\mathbf{x},t),\\
\mathbf{B}_{ji}(\mathbf{x},t)&:=&-\delta_j \mathbf{A}_i(\mathbf{x},t)+\delta_i \mathbf{A}_j(\mathbf{x},t).
\eea
with $\delta_i := \displaystyle{\frac{\delta}{\delta \mathbf{x}^i}}$. In above
$\underbrace{D_0\,\mathbf{x}^j\, \mathbf{B}_{ji}(\mathbf{x},t)}$ denotes the average over all possible insertions of
$D_0\, \mathbf{x}^j$ between $\mathbf{x}$'s of $\mathbf{B}_{ji}(\mathbf{x},t)$. The action (\ref{14}),
thanks to the symmetrization prescription, is invariant under the transformations below, in which
$U=\exp\big({\rm i} \Lambda(\mathbf{x},t)\big)$,
\bea\label{17}
\mathbf{x}&\to& \mathbf{x}'= U \, \mathbf{x}\, U^{-1},\nonumber\\
a_0(t) &\to&  a'_0(t)=U\, a_0(t)\, U^{-1} + {\rm i}\, U\, \frac{{\rm d}}{{\rm d}t}\, U^{-1},\nonumber\\
\mathbf{A}_i(\mathbf{x},t)&\to& \mathbf{A}'_i(\mathbf{x}',t)=
U\, \mathbf{A}_i(\mathbf{x},t)\,U^{-1}+{\rm i}\,U\, \delta_i\Lambda(\mathbf{x},t)\, U^{-1},\nonumber\\
\mathrm{A}_0(\mathbf{x},t)&\to&  \mathrm{A}'_0(\mathbf{x}',t)=
U \,\mathrm{A}_0(\mathbf{x},t)\,U^{-1}-{\rm i}\, U\partial_{\,t}\Lambda(\mathbf{x},t)\, U^{-1}.
\eea
Then the transformation rules for the field strengths are given by
\bea\label{18}
\mathbf{E}_i(\mathbf{x},t) &\to& \mathbf{E}\,'_i(\mathbf{x}',t)= U\, \mathbf{E}_i(\mathbf{x},t)\,U^{-1},\nonumber\\
\mathbf{B}_{ji}(\mathbf{x},t) &\to& \mathbf{B}\,'_{ji}(\mathbf{x}',t)= U\, \mathbf{B}_{ji}(\mathbf{x},t)\, U^{-1},
\eea
as they should. As $\mathbf{x}_i$'s in $\big(\mathrm{A}_0(\mathbf{x},t),\mathbf{A}(\mathbf{x},t)\big)$
just appear under the symmetrization prescription, one may wonder what would be the consequences of
adding terms with $\mathbf{x}_i\,$s' commutators. In the lowest order of $\mathbf{x}_i$'s and velocity,
the rotationally invariant action is given by
\bea\label{19}
S[a_0,\mathbf{x}]=\int {\rm d}t \;{\rm Tr} \hspace{-1.6em} &&\bigg(\frac{1}{2}\, m\, D_0\,\mathbf{x}\cdot D_0\,\mathbf{x}
- g \, \mathrm{A}_0(\mathbf{x},t) + g\, D_0\,\mathbf{x}\cdot \mathbf{A}(\mathbf{x},t)  \nonumber\\
&&\hspace{1em} -\frac{m}{4\ell^4}[\mathbf{x}^i,\mathbf{x}^j]\,[\mathbf{x}_i,\mathbf{x}_j]+\cdots \bigg),
\eea
in which $\ell$ is a constant with dimension of length. The action (\ref{19}) is known
to be the action of D0-branes of string theory, in the background of
(RR) gauge field $(A_0,\bm{A})$ \cite{myers}. From the string theory point of view, D0-branes
are point particles to which ends of strings are attached \cite{9510017}.
In a bound state of $N$ D0-branes, they are connected to each other by
strings stretched between them, and it can be shown that, by counting
the degrees of freedom for the oriented strings, the correct dynamical variables
describing the positions of D0-branes are $N\times N$ hermitian matrices \cite{9510135}.

As mentioned earlier, in \cite{fat-1}-\cite{fat-6} the consequences of introducing matrix coordinates
in a theory gauge were discussed, and here we do not intend to repeat these works.

We would like to mention a possible relation between the idea of the present work, and the observation
of \cite{jackiw} in the context of fluid mechanics of non-Abelian gauge theories. As discussed in
\cite{jackiw}, one may present formulations for a non-Abelian fluid, like a quark-gluon plasma,
either based on a so-called particle picture of matter, or on a field description of the fundamental
substratum. The interesting is that in the case with non-Abelian symmetry the two formulations
give different fluid equations. Accordingly one finds out that in the field based approach
the dynamical entities, such as velocity, possess group indices coming from the isospin structure
of the theory, very reminiscent of those appearing in this work.

As the final remark it would be worthwhile to mention another possible relation between
this work and the suggestion of a duality between space and the wave-function of quantum mechanics
\cite{xpsi}. In particular, it is observed that in the case of Dirac field, for which the wave-function
has extra indices of $spin(1,3)$, the proposed duality suggests that the relevant space coordinates
should be considered with extra structure, bringing them as matrices \cite{vancea}. It seems
quite expectable that the same line of \cite{vancea} would lead us that for the isospin
case one should expect to deal with matrix coordinates too.
\\
\\
\textbf{Acknowledgement}:  This work was partially supported by
the research council of the Alzahra University.


\end{document}